\documentclass[usenatbib,usegraphicx,letters]{mn2e}

\usepackage{times}
\usepackage{amssymb}

\def\<<{{\ll}}
\def\>>{{\gg}}

\def\spose#1{\hbox to 0pt{#1\hss}}
\def\ltwig{\mathrel{\spose{\lower 3pt\hbox{$\mathchar"218$}}
     \raise 2.0pt\hbox{$\mathchar"13C$}}}
\def\gtwig{\mathrel{\spose{\lower 3pt\hbox{$\mathchar"218$}}
     \raise 2.0pt\hbox{$\mathchar"13E$}}}

\newcommand{\beq}{\begin{equation}}
\newcommand{\eeq}{\end{equation}}
\newcommand{\beqa}{\begin{eqnarray}}
\newcommand{\eeqa}{\end{eqnarray}}
\newcommand{\aap}{A\&A}

\newcommand{\aj}{AJ}
\newcommand{\apj}{ApJ}
\newcommand{\apjl}{ApJ}

\newcommand{\mnras}{MNRAS}

\newcommand{\araa}{Ann.Rev.Ast.As}

\begin{document}
\title[X-ray Lightcurve of $\eta$ Carinae]{
Modeling the RXTE light curve of $\eta$~Carinae from a 
\\
3-D SPH simulation of its binary wind collision
% 3-D SPH simulation of colliding winds in $\eta$~Carinae
}

\author[Okazaki et al.]{
A.~T.~Okazaki$^{1}$\thanks{okazaki@elsa.hokkai-s-u.ac.jp}
,
S.~P.~Owocki$^{2}$\thanks{owocki@bartol.udel.edu}
,
C.~M.~Russell$^{2}$\thanks{crussell@udel.edu}
\& 
M.~F.~Corcoran$^{3,4}$\thanks{corcoran@barnegat.gsfc.nasa.gov} \\
\\
$1$ Faculty of Engineering, Hokkai-Gakuen University, 
Toyohira-ku, Sapporo 062-8605, Japan\\
$^2$ Department of Physics \& Astronomy,
University of Delaware, Newark, DE 19716, USA\\
$3$ Universities Space Research Association, 
7501 Forbes Blvd, 
Suite206, Seabrook, MD 20706, USA \\
$4$ Exploring the Universe Division, 
NASA Goddard Space Flight Center, Greenbelt, MD 20771 USA
}

\maketitle

\begin{abstract}
The very massive star system $\eta$~Carinae exhibits regular 5.54-year
(2024-day) period 
disruptive
events
in wavebands ranging from the radio to X-ray.
There is a growing consensus that these events likely stem from
periastron passage of an (as yet) unseen companion in a highly
eccentric ($\epsilon \sim 0.9$) orbit.
This paper presents three-dimensional (3-D) Smoothed Particle
Hydrodynamics (SPH) simulations of the orbital variation of the
binary wind-wind collision, and applies these to modeling the X-ray 
light curve observed
by the Rossi X-ray Timing Explorer (RXTE).
By providing a global 3-D model of the phase variation of
the density of the interacting winds,
% colliding wind density, 
the simulations allow computation of the
associated variation in 
X-ray absorption,
% the absorption of X-ray emission, 
presumed here to originate from near the apex of the wind-wind
interaction cone.
We find that the observed RXTE light curve can be readily fit if the observer's 
line of sight is within this cone along the general direction of apastron.
Specifically, the data are well fit by an assumed inclination $i = 45^{\circ}$ 
for the orbit's polar axis, which is thus consistent with orbital
% axis alignment 
angular momentum being 
along the inferred polar axis of the Homunculus nebula.
% However, 
% % though 
% inclinations in the broad range $i \approx 36^{\circ} - 90^{\circ}$ are also
% allowed.
The fits also constrain the position angle $\phi$ that an  orbital-plane 
projection makes with the apastron side of the semi-major axis,
strongly excluding positions $\phi < 9^{\circ}$ along or to the retrograde
side of the axis,
% % but allowing a range $\phi \approx 18^{\circ} - 45^{\circ}$
% % of prograde positions, 
with the best fit position given by $\phi = 27^{\circ}$.
% and $i=45^{\circ}$.
Overall the results demonstrate the utility of a fully 3-D dynamical
model for constraining the geometric and physical properties of this
complex colliding-wind binary system.

\end{abstract}


\section{Introduction}
 \label{s_intro}
$\eta$~Carinae is one of most  remarkable star systems in the galaxy.
Its extreme luminosity, estimated today at some
$5 \times 10^{6} L_{\odot}$, implies a massive ($M> 100 M_{\odot}$)
primary star very close to the Eddington limit.
% It is one 
One of the most extreme examples of the class of Luminous Blue Variable 
(LBV) stars,
its historical light curve shows irregular
brightenings, the greatest of which occurred in the 1840's, when its
luminosity is  estimated to have approached $25 \times 10^{6} L_{\odot}$.
This was
accompanied by the ejection of some 10-20~$M_{\odot}$, forming
what is seen today  as the bipolar Homunculus nebula.
In general, $\eta$~Carinae is a key object in our understanding of the formation
and evolution of extremely massive stars 
\citep[see, e.g.,][]{DH97}.
% (see, e.g., Davidson \& Humphreys 1997).

\begin{figure}
\begin{center}
    \includegraphics[width=8.1cm]{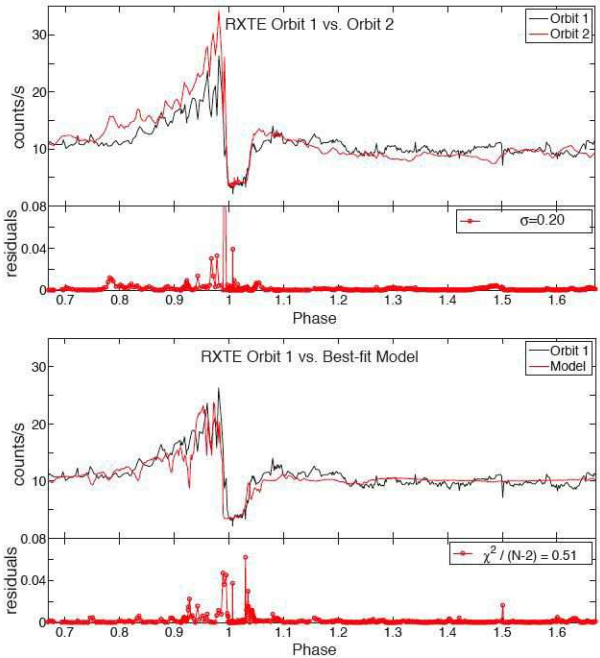}
\caption{
{\em Top:} 
RXTE light curve vs.\ orbital phase, comparing data from the first
(black) and second (red) orbital periods.
% with the lower box showing
% the square residuals, normalized by the total deviation $N \sigma^{2}$
% defined in eqn.\ (\ref{sigobs}), which results in $\sigma = 0.20$.
{\em Bottom:}
Comparison of the initial orbit RXTE light-curve (black) with our
best-fit model (red), with observer position angles $i=45^{\circ}$ and
$\phi = 27^{\circ}$, as marked by ``X'' 
in the $\chi^{2}$ contour plots in figure \ref{fig:chisq}.
For both comparisons, the lower boxes plot
the square residuals 
% [(model/observed-1)$^{2}$]
of the fit,
% for all $N=481$ first-orbit RXTE datapoints,
normalized so that the sum equals 1.
% To highlight the level of deviation, a cyan line is drawn between the red and 
% black curves for each RXTE-Orbit1 data point.
% The green curve added to the lower panel emphasizes that a viewing from
% $180^{\circ}$ opposite to the best-fit perspective provides a very poor fit
% to the data.
% representing 100\% of the reduced $\chi^{2}$ for each model.
% The light gray line shows the first cycle of RXTE data shifted by one period.
% Taken from \citet{rus08}.
}
\end{center}
\label{fig:lightcurves}
\label{fig:rxte}
% \vskip -0.5cm
\end{figure}

An important advance in the observational study of $\eta$~Carinae came
from the identification of periodic, near-IR variations 
\citep{w94, dam96}
% (Damineli 1996; Whitelock et al. 1994) 
that are stable over 
many decades, along  with correlated variability in the
radio \citep{dun95}
% (Duncan et al. 1995) 
and X-ray  \citep{cor95} wavebands.
% (Corcoran et al. 1995) 
The variability is especially dramatic in the 2-10~keV 
X-ray band, where the spatially unresolved X-ray flux drops by 
about a factor of 100 for 3 months, as shown by daily monitoring
with the  Rossi X-Ray Timing Explorer (RXTE) during the 
X-ray minimum of 1997-1998 
\citep{ish99},
% (Ishibashi et al. 1999), 
and again during the 2003 X-ray minimum 
\citep{ham07}.
% (Hamaguchi et al. 2007).
% The top panel in figure \ref{fig:lightcurves} compares the RXTE light curve 
% vs.\ phase for the first (red) and second (blue) observed orbital periods,
% centered respectively around the 1998 and 2003.5 minima
The top panel of figure \ref{fig:rxte} compares the RXTE lightcurve vs.\ phase 
over the two full orbital cycles for 1996-2001 and 2002-2007 \citep{cor05}.

Analysis of this X-ray emission and light curve has provided important 
clues about the likely general nature of the system.
First, the relative hardness of the X-rays suggests they must originate from
the post-shock regions of a relatively fast wind 
($\sim 3000 \, {\rm km\,s^{-1}}$) from an otherwise unseen 
% (or not yet seen) 
companion star,
confined by the much denser, but slower 
($\sim 500-800\, {\rm km\,s^{-1}}$) wind from the primary
\citep{pitt98, cor01, pc02}.
% (Pittard et al. 1998; Ishibashi et al. 1999;  Corcoran et al. 2001a; 
% Pittard \& Corcoran 2002).
The sharpness of the 
ingress and egress
% X-ray minimum 
suggests moreover that the X-ray
emission source must be relatively compact, probably originating
mostly just inside the 
% head
stagnation point
of the wind-wind shock cone, along the line
between the stars.
And given the very high density of the primary wind, the detection of
X-rays during most of the period suggests an observer perspective that
looks through a relatively transparent cavity carved out by the
relatively low-density  secondary wind \citep{cor05}.
% WHAT IS BEST REFERENCE FOR THIS?

% But a
A key hindrance to moving beyond this general picture has been the 
% inherent difficulty of computing 
lack of a 3-D hydrodynamical wind-interaction model that fully
accounts for the orbital motion,
which can be especially important for the sharp variations
near periastron.
The present paper applies Smoothed Particle Hydrodynamics (SPH)
simulations to provide such a 3-D model 
% of the wind-wind collision
throughout the full elliptical orbit of the binary components
\citep{oka08}.
For simplicity, the initial simulations here assume {\em isothermal} 
flow with a fixed common temperature for both winds.
As such they do not directly model the shock-heated gas that is the 
cause of the X-ray {\em emission}.
But the simulations do provide a fully 3-D, time-dependent description 
of the relatively cool material that is the source of X-ray 
{\em absorption}.
By assuming a simple point-source model for the X-ray emission,
located just within the head of the wind-wind shock interaction front
(see figure~\ref{fig:angles}), 
the model allows computation of the phase-variable X-ray attenuation, 
and thus X-ray light curve, 
for any assumed observer position.
As detailed below, with quite nominal binary wind parameters adopted
from previous analyses, the overall model, once adjusted to an optimal
viewing angle, reproduces the observed RXTE light curve remarkably well
(see figure~\ref{fig:lightcurves}).

\begin{table}
  \begin{center}
  \caption{Stellar, wind, and orbital parameters}
  \label{tbl:params}
  \begin{tabular}{lcc}
  \hline
  \multicolumn{1}{c}{Parameters}
  & $\eta$~Car A & $\eta$~Car B \\
  \hline
  Mass ($M_{\odot}$) & 90 & 30 \\
  Radius ($R_{\odot}$) & 90 & 30 \\
  Mass loss rate ($M_{\odot}\,{\rm yr}^{-1}$) & $2.5 \times 10^{-4}$ & $10^{-5}$ \\
  Wind velocity (${\rm km\,s}^{-1}$) & 500 & 3,000 \\
  Wind temperature (K) & $3.5 \times 10^{4}$ & $3.5 \times 10^{4}$ \\
  \hline
  Orbital period $P$ (d) & \multicolumn{2}{c}{2,024} \\
  Orbital eccentricity $e$ & \multicolumn{2}{c}{0.9} \\
%   Semi-major axis $a$ (cm) & \multicolumn{2}{c}{$2.3 \times 10^{14}$} \\
  Semi-major axis $a$ (AU) & \multicolumn{2}{c}{$15.4$} \\
  \hline
  \end{tabular}
  \end{center}
\vskip -0.1cm
\end{table}

\section{Model Specifications}

The simulations presented here were performed with a 3-D 
% Smoothed Particle Hydrodynamics (SPH) 
SPH code based on a version originally
developed by 
% \citet{ben90a}, 
\citet{ben90b}
and \citet{bat95}.  
Using a variable smoothing length, the SPH equations with the standard
cubic-spline kernel are integrated with individual time steps for each
particle.  
In the implementation here, the artificial viscosity
parameters are  $\alpha_{\rm SPH}=1$ and $\beta_{\rm SPH}=2$.

The two winds are modeled by an ensemble of gas particles that are 
continuously ejected with a given outward velocity at a radius just
outside each star, coasting from there without any net external forces, 
effectively assuming that gravity is canceled by radiative driving terms. 
Perhaps more significantly, the simulations also assume both winds to be 
isothermal, with a common ``warm'' temperature.
% \footnote{
(The specific
temperature, 
set to be 
comparable to the stellar
% a typical hot-star 
effective temperature
$T = 35,000$~K, 
% is rather arbitrary, since within a broad range at this
% magnitude, the temperature 
has little effect on the flow dynamics or X-ray absorption.)
% }.
% near the effective temperature of the primary($T = 35,000$~K).
% $T= 3.5 \times 10^{4}$~K.
This is a serious simplification, made to bypass the need to resolve
the complex cooling regions near the wind shocks, which is generally
difficult in a 3-D model, particularly with an inherently viscous
method like SPH.
While this does allow a quite realistic account for the 3-D absorption
by radiatively cooled material, it means that the expected X-ray
emission from shock heating must be added separately
(see \S 4).

In a standard $xyz$ Cartesian coordinate system, we set the binary orbit 
in the $x$-$y$ plane, with the origin at the system centre of mass,
and
semi-major axis along the x-axis
% apastron of the major axis in the $+x$-direction
(see figure~\ref{fig:angles}).
% The outer simulation boundary is set at a radial distance $r$ from the
% origin, with either $r=10.5a$  or $r = 105a$,
% where $a$ is the semi-major axis of the binary orbit. 
The outer simulation boundary is set at a radial distance $r=10.5a$ from 
the origin, where $a$ is the semi-major axis of the binary orbit. 
Particles crossing this boundary are removed from the simulation. 
By convention, we define $t=0$ (and zero phase) to be at periastron
passage.
Table~\ref{tbl:params} summarizes the stellar, wind, and orbital
parameters, largely adopted from those derived previously by
\citet{cor01} and \citet{hil01}.
A key parameter for the global form of the wind interaction is the
ratio $\eta$ of {\em wind momentum} ${\dot M} v$ between
the primary to secondary wind, 
which here has a value $\eta \approx 4.2$.
% In the following, $t=0$ (Phase 0) corresponds to the periastron passage.
% 
% A key parameter for the global form of the wind interaction is the
% momentum ratio of the primary to secondary wind, which here has a
% value $\eta \approx 4.2$.
Simple ram pressure balance then implies that, for a binary
separation $D$, the interface should be located at a distance
$d = D/(1+\sqrt{\eta}) \approx 0.33 \, D$ from the secondary star
\citep{SBP92, CRW96}.
% The observed hard X-ray emission originates from the shocked secondary
% wind just inside the head of the full interface cone that wraps around
% the secondary star.

\begin{figure*}
\begin{center}
% \resizebox{\hsize}{!}{
    \includegraphics[width=17cm]{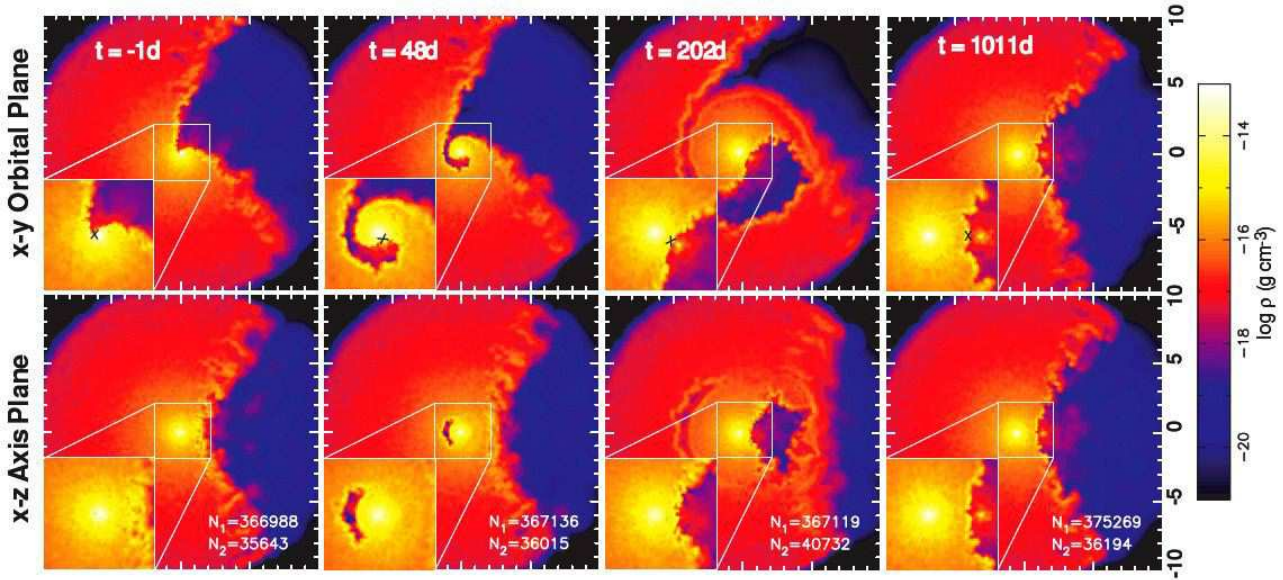}
% }
\caption{
Snapshots of 3-D SPH simulation of wind-wind collision 
at times that, from left to right, are -1, +48, +202, and +1,011 days from
periastron.
The color scale shows the density
(on a logarithmic scale with cgs units)
in the $x$-$y$ orbital plane (top) and in the $x$-$z$  perpendicular plane
containing the orbital and major axes (bottom).
The main figures are for a square region $\pm 10 a$ about the
system centre of mass, while the lower-left insets show a factor two
magnification of the inner $\pm 2 a$;
this shows more clearly the
interaction front where the X-ray source is assumed located, marked by
an X between the brighter and dimmer spots that represent the
primary and secondary stars, within the apex of the lower density wind.
Annotations give the time (in days) from periastron
passage and the number of particles ($N_{1}, N_{2}$) in the primary and
secondary winds.
% wind, $N_{1}$, and in the secondary wind, $N_{2}$.
}
\end{center}
% \vskip -0.5cm
\label{fig:sphden}
\end{figure*}

\begin{figure}
\begin{center}
    \includegraphics[width=8.0cm]{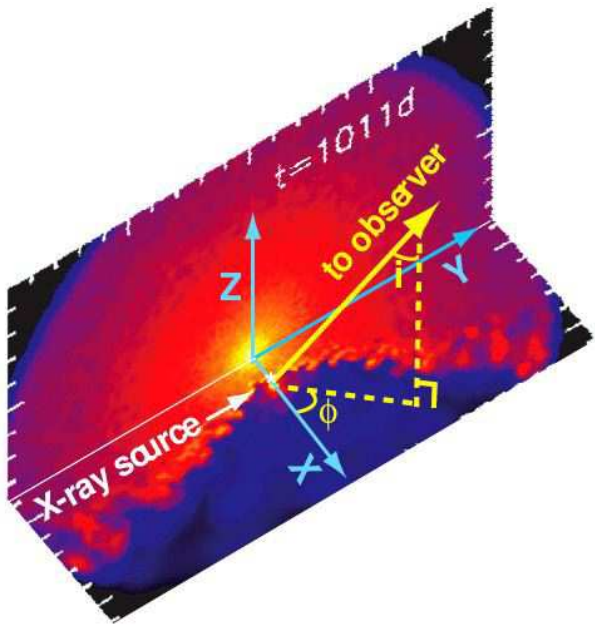}
\caption{
Schematic to illustrate observer position defined by 
inclination angle $i$ to the orbital axis $z$, and by
equatorial projection angle $\phi$ relative to major axis $x$.
The background orthogonal planes represent slices of density from our
3-D SPH simulation, shown here near apastron ($t=1011$~d)
for both the $x$-$y$ orbital plane
and the $y$-$z$ plane orthogal to the major axis. 
The X-ray source is assumed to be just inside the head of the bow front, 
at a distance $0.25 D$ from the secondary, where $D$ is the current 
binary separation.
The observer direction shown is roughly consistent with that inferred
by the best-fit model to the RXTE lightcurve.
% Note that the azimuthal position $\phi$ is related to the ``argument of periapsis",
% Ê$\omega$, defined by the angleÊbetween the Êperiastron side of the major axis and
% Êthe intersection between the orbital plane and a reference plane.
% Choosing that reference to be the plane of the sky, we find $\omega= 270^{\circ} - \phi$.
Note that, in the conventional notation of binary orbits,
using a reference plane perpendicular to the line of sight,
the  ``argument of periapsis'' $\omega= 270^{\circ} - \phi$.
}
\end{center}
% \vskip -0.3cm
\label{fig:angles}
\end{figure}

% \section{Structure and phase variation of colliding winds}
\section{Phase variation of colliding winds}

% For the simulation covering radii $r \le 10.5a$ from the centre of mass,
% Figure~\ref{fig:sphmodel} 
Figure~\ref{fig:sphden} illustrates SPH simulation results 
for the density 
at 4 phases from near periastron (left) to near apastron (right), 
plotted in both the orbital plane ($x$-$y$; top row), and the
perpendicular plane through the orbital and major axes 
($x$-$z$; bottom row).
% The colorscale represents the density in cgs units on a logarithmic scale.
% From top down, the rows show 4 snaphots with increasing orbital phase:
% near periastron ($t = -1$~d);
% during  X-ray minimum ($t = 48$~d);
% after the X-ray minimum ($t = 202$~d);
% and finally near apastron ($t= 1011$d).
% % The numbers give the instantaneous number of active SPH particles from the 
% % primary ($N_{1}$) and secondary $(N_{2})$.
% The insets provide a factor 2 magnification of the inner $\pm 2a$
% about the centre of mass.
% The bright spot near the origin represents the primary ($\eta$~Car A), while
% the smaller bright spot close to the apex of the lower density wind represents 
% the secondary ($\eta$~Car B).
% 
Although instabilities in the wind-wind interaction lead to substantial 
stochastic variations and clumping, 
% the global structure is still easily traced. 
% In particular, 
one can still see quite vividly how the lower-density, faster wind 
from the secondary carves out a cavity in the higher-density, 
slower wind from the primary.
% As expected, the shape of the collision interface around apastron, 
% where the orbital speed  of the secondary is only 
% $\sim 20\,{\rm km\,s}^{-1}$ with respect to the primary,
% is in agreement with the analytical one \citep[e.g.,][]{ant04}. 
Throughout most of the period centered around apastron, this cavity has
a relatively simple, 2-D axisymmetric, conical form similar to the
apastron snapshot at $t=1011$~d,
with a fixed opening half-angle $\alpha \approx 60^{\circ}$.
This is in good agreement with 2-D analytic \citep{CRW96}
% (Canto et al. 1996)
% Canto, J., Raga, A. C., & Wilkin, F. P. 1996, ApJ, 469, 729
and numerical models \citep{pc02}
% (Pittard \& Corcoron 2002) 
that ignore orbital
motion, which near apastron is indeed small (ca.\ $~20~{\rm km\,s}^{-1}$) 
compared to the flow  speed of either wind.
% ($> 500~{\rm km\,s}^{-1}$).

But near periastron, the faster variation, closer separation, and
higher orbital speed (up to $\sim 360~{\rm km\,s}^{-1}$) 
all work to 
distort the structure.
% make the structure fully 3-D.
In the approach up to periastron, the 2-D interface first starts to bend, 
but then, as the secondary whips around the opposite side of the
primary, the secondary wind cavity becomes fully enshrouded by the
denser, primary wind.
Over time, the segment of this shell expanding toward apastron
dissipates, and the nearly 
2-D
axisymmetric 
structure is again recovered.

\begin{figure*}
% \begin{center}
  \includegraphics[width=17.5cm]{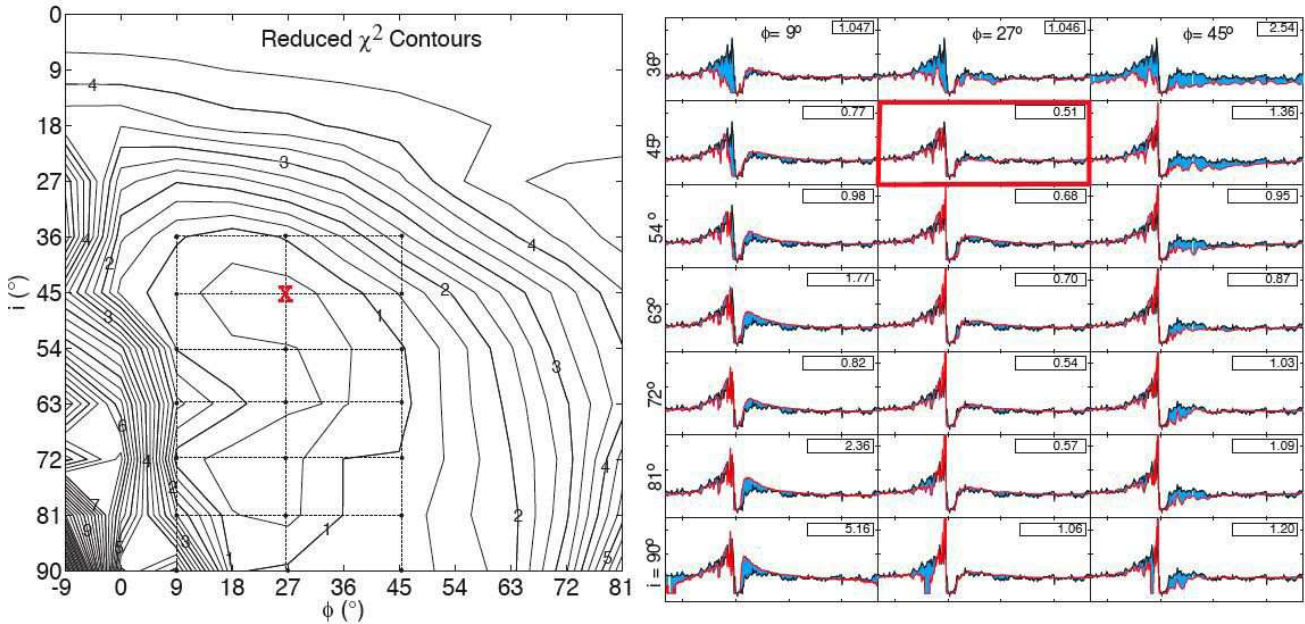}
\caption{
{\em Left:} 
Contours of reduced chi-squared, $\chi^{2}/(N-2)$,
for deviations between RXTE lightcurve for orbit~1 
and models with observer inclination angle $i$ 
and prograde angle $\phi$,
plotted for a model grid with
$9^{\circ}$ increments in both $i$ and $\phi$.
The large ``X'' denotes the best-fit model
plotted in figure~\ref{fig:lightcurves}.
{\em Right:}
Mosaic of lightcurves for inclinations and azimuths denoted by dots
within the box bracketing the broad $\chi^{2}$ valley on left.
The curves compare model (red) vs.\ orbit~1 RXTE data (black),
with the difference shaded in cyan.
The box in each panel gives the reduced chi-squared value.
The red box highlights the best-fit model,
with $i=45^{\circ}$, $\phi=27^{\circ}$, and $\chi^{2}/(N-2) = 0.51$.
}
% \end{center}
\vskip -0.0cm
\label{fig:chisq}
% \label{fig:mosaic}
\end{figure*}

\section{Modeling the RXTE light curve for $\eta$~Car}

To illustrate the diagnostic potential of this 3-D SPH simulation, 
% let us 
we
now use it to model the X-ray light curve observed by RXTE.
The solid black curve in figure \ref{fig:rxte} shows 
this light curve
% the RXTE light curve \citep{cor05},
% Corcoran 2005, AJ 129,2018 - 2025 from 1996 to 2004, and 
% for the years 1996-2004, covering both the 1998 and the 2003 minima.
% While the sharpness of the drops to this minima
for the years 1996-2007, covering the two initial full periods
spanning both the 1998 and 2003.5 minima \citep{cor05}.
While the sharpness of the drop to these minima
seems suggestive of an eclipse-like event, 
the overall asymmetry does not fit the normal form of a
stellar eclipse.
The pre-event rise can likely be attributed to the $1/D$ scaling of the
shock emission with the declining binary separation distance $D$
toward periastron. 
But it is been a subject of debate whether the
sharp drop and lack of a symmetric post-event peak reflects some kind
of quenching of the X-ray emission \citep{ham07},
% (REFERENCE NEEDED HERE),
or is mainly just due to variations in X-ray absorption.

To explore the latter possibility, 
% let us 
we
combine the variable absorption
column derived from the SPH simulations with a simple point-source
model for the X-ray emission.
The strongest X-ray emission is expected to come from the shock of the faster 
secondary wind in the region just within the head of the wind-wind 
interaction front.
In terms of the binary separation $D$, 
that interaction front is itself a distance
$d=D/(1+\sqrt{\eta}) = 0.33 \, D$ from the secondary. 
Our model thus assumes a point source of X-ray emission located along 
the line of separation at a fixed fractional distance from the secondary,
% chosen here as $d_{x} = 0.25 \, D$.
given by $d_{x} = f_{x} \, D$, where $f_{x} < 0.33$.
% % We find a good fit for $f_{x} = 0.25$, but in \S 4 we also
% % explore cases with $f_{x}=0.3$ and $f_{x}=0.2$.
% % We find a good fit for $f_{x} = 0.25$, and so, though we have also
% % explored models with $f_{x}=0.3$ and $f_{x}=0.2$, we concentrate
% % discussion here on this more central case.
% Though we have also explored models with $f_{x}=0.2$ to $f_{x}=0.3$,
% we focus the discussion here specifically on the intermediate 
% case with $f_{x} = 0.25$.
We have explored models with $f_{x}=$~0.20, 0.25, and 0.30, but since
the results are all qualitatively similar, we focus 
% the specific discussions 
here just on the intermediate case with $f_{x} = 0.25$.

Following the expected scaling for emission by adiabatic shocks in
wind-wind collisions \citep{SBP92, pc02},
% (REF), 
we assume the phase variation of the
X-ray source brightness varies with the inverse of the current
stellar separation, $L_{x} \sim 1/D$.
Defining then the time-variable mass column depth from the X-ray source to 
the observer as $m_{o} (t)$, the model X-ray light curve takes the
form,
\beq
L_{x,mod} (t) = \frac{A}{D(t)} e^{-\kappa m_{o} (t)} + B 
\, ,
\label{lxmod}
\eeq
where $A$ and $B$ are normalization constants fixed to match the
observed X-ray counts respectively at apastron and post-periastron
minimum.
Assuming a characteristic bound-free opacity
$\kappa \approx 5 \, {\rm cm^{2} \, g^{-1}}$ 
\citep[see, e.g., fig. 5 of][]{ant04}
for the relevant RXTE 
energy band (2-10~keV), we then compute the phase variation of
absorption from this X-ray source to trial observers over wide range of
position angles.
As illustrated in figure~\ref{fig:angles}, this observer position is defined
by the inclination $i$ to the orbital axis, and by an orbital plane projection
that makes a prograde direction angle $\phi $ with the $+x$ axis direction 
toward apastron.
% \footnote{
% This azimuthal position is related to the "argument of periapsis",
% Ê$\omega$, defined by the angleÊbetween the Êperiastron side of the major axis and
% Êthe intersection between the orbital plane and a reference plane.
% Choosing that reference to be the plane of the sky, we find $\omega= 270^{\circ} - \phi$.
% }.

\vskip -0.5cm
\section{Varying Observer Position for Best Fit}
% lightcurve}

We have computed a grid of model X-ray light curves $L_{x,mod}(t)$ vs.
temporal phase $t$ for a full range of 
observer's position angles
% \footnote{The north-south symmetry makes inclinations
% $90^{\circ}<i<180^{\circ}$ redundant.}
% $0 < i < 90^{\circ}$ and $0 < \phi < 360^{\circ}$, 
$i$ and $\phi$,
varying both in increments of $9^{\circ}$.
The lower panel of figure~\ref{fig:lightcurves} 
% plot the RXTE lightcurve (red) over the first orbital period.
% The top panel compares this the RXTE data from the second orbit,
% showing that the variation, while clearly periodic, has significant cycle 
% to cycle differences.
% The middle and lower panels 
compares the first-orbit RXTE light curve
with the resulting best-fit model, 
for which 
% observer position angles 
$i=45^{\circ}$ and $\phi=27^{\circ}$.
The agreement is as good or somewhat better than the
internal agreement between the first and second orbit cycles of RXTE
observations, as shown by the black vs.\ red curves in the upper panel.
% In each case, the lower boxes plot the square residuals for each
% observed time phase $t_{i}$, normalized to sum to unity.
% by dividing by the total variation $N \chi_{sq}$ for each model.
% 
In fact, the random variations in model X-rays during the general
rise before periastron appear to be statistically quite similar to 
RXTE variations during this phase, though of course the random nature means
they don't match in detail.
In the model, these variations arise entirely from changes in
absorption due to clumping in the wind interaction region of the SPH
simulation, suggesting then that the observed variation might likewise be
due to clump absorptions rather than, e.g., 
% ``flares'' in X-emission
increases in the temperature or emission measure of 
the shock X-ray emitting region
\citep[cf.][]{ham07}.

% Qualitatively, it is clear that our simple model can give a good overall 
% agreement with the observed lightcurve, but let us now analyze the
% fit statistics quantitatively to estimate statistical goodness
% of this best fit, along with the range of alternative acceptable 
% observer positions.

% By comparison with the observed RXTE counts $L_{x,obs} (t_{i})$ 
% for each of the $N=481$ data times $t_{i}$ spanning the
% first full 2024-day orbit observed with RXTE,
% we compute 

For each observer position we quantify the level of agreement with the
first RXTE orbit by 
the usual
% $\chi^{2}$ 
statistical measure of merit,
\beq
\chi^{2} = 
% \frac{1}{N \sigma^{2}} 
\frac{1}{\sigma^{2}} 
\sum_{i} 
\left [ \frac{ L_{x,mod} (t_{i})}{L_{x,obs} (t_{i})} - 1 \right]^{2}
\, ,
\label{chisqdef}
\eeq
where we have assumed the data can all be characterized by a common 
{\em fractional} mean-square deviation $\sigma^{2}$.
% For the RXTE data, we assume this to be dominated by apparently random
% variations in the source emission, e.g. due to wind clumping, with
% relatively important contribution from actual measurment error.
For the RXTE data, the contribution from measurment error is 
relatively unimportant compared with the inherent, apparently random
variations in the observed X-rays, e.g. perhaps due to wind clumping.
Moreover, in figure \ref{fig:rxte} the comparison of the
RXTE light curves for successive orbital periods
shows a systematic change, indicating a cycle-to-cycle variation that
% would not be 
is not
accounted for in our basic model.
% As such, 
We 
thus
estimate the inherent deviation by computing the
averge mean-square deviation between each of these first two observation cycles,
\beq
\sigma^{2} \approx \frac{1}{N} \sum_{i} 
\left [ \frac{ L_{x,obs} (t-P)_{i}}{L_{x,obs} (t_{i}) } - 1 \right]^{2}
\, ,
\label{sigobs}
\eeq
where $L_{x,obs} (t-P)_{i}$ represents data from the second orbit
shifted back by one period $P = 2024$~d and interpolated onto the data
times of the first cycle.
Application of this procedure for the RXTE data yields an 
estimated relative rms error, $\sigma = 0.20$.

% % % % % 

Figure~\ref{fig:chisq} plots contours of 
the {\em reduced} chi-squared, $\chi^{2}_{red} \equiv \chi^{2}/(N-2)$,
for the most relevant subset of our model grid, with 
% inclination ranging  from orbital axis to plane ($0^{\circ} < i < 90^{\circ}$), and
azimuth spanning a $90^{\circ}$ range  from just retrograde to strongly prograde 
of the major axis ($ -9^{\circ} < \phi < 81^{\circ}$).
Noting the overall north-south symmetry, the inclination spans the
full range of just the northern hemisphere, $0 < \phi < 90^{\circ}$.
The formal best-fit model, marked with an ``X'', has observer
position angles $i=45^{\circ}$ and $\phi = +27^{\circ}$,
with a $\chi^{2}_{red} = 0.51$ that is quite significantly {\em below}
the unit value required for a good fit.
(This suggests 
our derived $\sigma=0.2$ may be about a factor $\sqrt{2}$ 
overestimate.)
% of the true rms error.
% The light curve and residuals for this best-fit position are plotted in the
% lower panel of figure~\ref{fig:lightcurves}.

The contours also help identify the allowed range of
viewing angles, though this can be difficult to quantify rigorously.
A common approach \citep{ptvf07} is to define the difference in chi-squared
relative to the best-fit model, which here gives
\begin{equation}
    \Delta \chi^{2} \equiv \chi^{2} (i,\phi) - \chi^{2}_{min}
    = 479 \times (\chi^{2}_{red}-0.51)
    \, ,
\label{delchisq}
\eeq
where $479=N-2$ represents the number ($N=481$) of data points in the first 
RXTE orbit, minus the two degrees of freedom ($i,\phi$) in the data
fit.
It turns out even models neighboring the best-fit have 
$\Delta \chi^{2} > 10$, sometimes several tens or even in the
hundreds; 
by formal statistics they would all
be excluded at well above the 99\% confidence level.
Taken at face value, this implies that, around the best-fit values
$i=45^{\circ}$ and $\phi=27^{\circ}$, the range in both allowed viewing angles 
is less than the $\pm 9^{\circ}$ of the model grid.
But this approach very likely greatly overstates the real exclusion
probability, given that we are fixing many model parameters about the 
orbit, winds, location of the X-ray source, etc.

Nonetheless, even the reduced chi-squared contours do seem to strongly exclude
azimuths with $\phi < 9^{\circ}$ that are near or retrograde of the major 
axis;
likewise, inclinations near the orbital axis, i.e. with $i \le
36^{\circ}$, seem also excluded.
On the other hand, viewing angles over the broad plateau within
$\chi ^{2}_{red} \approx 1$ (representing a doubling of the minimum)
might still be allowed.
The range in azimuth  ($ 9^{\circ} < \phi < 45^{\circ}$)
and inclination ($ 36^{\circ} < i < 90^{\circ}$)
essentially just places the observer on the prograde side within the wind
interaction cone of half-angle $\sim 60^{\circ}$ about the apastron side
of the major axis $x$ (see figure~\ref{fig:angles}).

The right panel of figure~\ref{fig:chisq} compares lightcurves for 
viewing angles that bracket this allowed region.
Comparison of the relative area of cyan shading between the observed
% (black)
and model 
% (red) 
curves supports the view that the full range of inclination
$i=45^{\circ}-81^{\circ}$ around $\phi = 27^{\circ} \pm 9 ^{\circ}$ 
give an acceptably good fit,
while models outside this range do not.

% i.e. within a prograde range of 
% azimuth  ($ 18^{\circ} < \phi < 45^{\circ}$), 
% and a quite broad range of  inclination ($ 36^{\circ} < i < 90^{\circ}$)
% extending to the orbital plane.
% Note in fact that there are two distinct islands of minimum $\chi^{2}$,
% centreed on the two observer positions marked by $A$ and $B$, with 
% distinct inclinations $i=45^{\circ}$ and $i=72^{\circ}$, but the same
% prograde azimuthal position $\phi=18$.
% The light curve and residuals for the best-fit position are plotted in the
% lower panel of figure~\ref{fig:lightcurves}.

\section{Conclusions and future work}

The relative ease and natural way that the observed RXTE light curve is 
fit by this 3-D absorption plus point-source-emission model provides good
evidence for the basic validity of the overall paradigm, the key
features of which are:
\begin{enumerate}
\item{} A highly elliptical orbit with the observer viewing from the
general direction of apastron and prograde of the semi-major axis, through a
cavity carved out in the slower, denser, primary wind by the faster,
less-dense, secondary wind.
\item{} A relatively localized X-ray source located on the secondary side 
of the interaction front between the stars.
\item{} The X-ray mininum arising from a ``wind eclipse'' of this
localized source as the primary wind engulfs the secondary wind just after
periastron.
\end{enumerate}

Note that the last point implies that ``quenching'' of the X-ray
emission is not likely to be a dominant effect in causing the
broad-band X-ray minimum.
On the other hand, recent analyses \citep{ham07} suggest that some sort of 
spectral variation of emission may be necessary to explain observed changes 
in the X-ray hardness.
In future work, we plan to extend our analyses to include more
realistic models of the energy dependence of both
the emission and absorption, with a particular 
focus on explaining such spectral energy and hardness variations.

% We have used  3-D SPH simulations to model the wind collision interaction 
% in the massive binary $\eta$~Carinae.
% The results provide vivid illustration of how the lower density,
% faster wind from the secondary ($\eta$~Car B) carves out a cavity in
% the higher-density, slower wind from the primary ($\eta$~Car A).
% With an optimal viewing angle of $i = 54^\circ$ and $\phi = 36^\circ$,
% where $i$ is the inclination angle and $\phi$ is the azimuth measured 
% from apastron in the prograde direction, the model gives an excellent fit
% with the RXTE X-ray light curve
% \citep{rus08}.

% \acknowledgments

% \section{Acknowledgements}

\vskip 0.25cm
\noindent
{\bf Acknowledgements}

\noindent
% \begin{acknowlegements}
A.T.O. thanks the Japan Society for the Promotion of Science for
financial support via Grant-in-Aid for Scientific Research
(16540218).
SPH simulations were performed on HITACHI SR11000 at Hokkaido
University Information Initiative Center.
S.P.O.\ acknowledges support of NSF grant AST-0507581 and NASA
grant Chandra/TM7-8002X.
C.M.R.\ acknowledges support of a NSF GK-12 fellowship.
MFC acknowledges support from NASA and the RXTE program.
We thank D. Cohen for many helpful comments.
% and discussons.
% \end{acknowlegements|}

\end{document}